\documentclass{article}
\usepackage{spconf,amsmath,graphicx,hyperref,amssymb,enumitem,tabularx,booktabs,pifont,tikz,pgfplots,xcolor}
\usepgfplotslibrary{groupplots}
\definecolor{mygreen}{RGB}{124, 205, 124}
\definecolor{myred}{RGB}{238, 162, 173}
\definecolor{mygray}{gray}{0.8}

\title{Multi-Agent Honeypot-Based Request-Response Context Dataset for Improved SQL Injection Detection Performance}
\name{
\begin{tabular}{c}
Hao Yu$^{1}$, Hui Li$^{1*}$, FengYuan Shi$^{1}$, Wenjie Yu$^{1}$, PinHan Ho$^{2}$, Zehua Wang$^{3}$, Bin Wang$^{1*}$\thanks{*Corresponding author. This work is supported by Guangdong Provincial Laboratory of Ultra High Defnition Immersive Media Technology (GrantNo.2024B1212010006)}
\end{tabular}
  }
\address{$^{1}$School of Electronic and Computer Engineering, Peking University\\
$^{2}$Department of Electrical and Computer Engineering, The University of Waterloo\\
$^{3}$Department of Electrical and Computer Engineering, The University of British Columbia}
\begin{document}
%
\maketitle
\begin{abstract}

SQL injection remains a major threat to web applications, as existing defenses often fail against obfuscation and evolving attacks because of neglecting the request-response context. This paper presents a context-enriched SQL injection detection framework, focusing on constructing a high-quality request-response dataset via a multi-agent honeypot system: the Request Generator Agent produces diverse malicious/benign requests, the Database Response Agent mediates interactions to ensure authentic responses while protecting production data, and the Traffic Monitor pairs requests with responses, assigns labels, and cleans data, yielding totally 140,973 labeled pairs with contextual cues absent in payload-only data. Experiments show that models trained on this context dataset outperform payload-only counterparts: CNN and BiLSTM achieve over 40\% accuracy improvement in different tasks, validating that the request-response context enhances the detection of evolving and obfuscated attacks.
\end{abstract}
\section{Introduction}
\label{sec:intro}

SQL injection (SQLi) remains one of the most critical threats to web applications, enabling adversaries to bypass authentication, exfiltrate sensitive data, or gain persistent control of back-end systems \cite{crespo2023sql}. Despite long-standing mitigation techniques such as input validation \cite{aliero2020algorithm}, parameterized queries \cite{roy2022sql}, and Web Application Firewalls (WAFs) \cite{thalji2023ae}, SQLi continues to rank among the OWASP Top Ten vulnerabilities \cite{nasereddin2023systematic}. A common shortcoming of prevailing defenses is their focus on isolated inputs or pre-defined signatures, neglecting the semantic alignment between user requests and server responses that often reveals attack success, such as database error messages or anomalous outputs. As a result, these defenses are increasingly insufficient against obfuscated attacks, dynamic query construction, and rapidly evolving attack vectors.


Recent advances in machine learning and large language models \cite{wang2025reflexgen} have significantly improved SQLi detection by capturing payload semantics, outperforming rule-based systems and signature-driven defenses \cite{lu2023semantic}. However, most ML-based detectors still suffer from a critical flaw: they prioritize the analysis of isolated payloads while neglecting the bidirectional context of HTTP request-response pairs \cite{gu2019diava}. This oversight means they cannot fully exploit critical signals like response latency, status codes, or server-generated error text, all of which are essential for distinguishing successful SQLi from benign queries or failed attack attempts, and for generalizing to unseen obfuscated attack techniques.

To address this context gap, we build a high-quality request-response dataset via a multi-agent honeypot system: the Request Generator Agent generates diverse malicious and benign traffic to simulate real-world attack scenarios, the Database Response Agent mediates interactions to ensure authentic server responses while safeguarding production data through shadow database isolation, and the Traffic Monitor Agent pairs requests with responses, assigns precise labels, and cleans data—ultimately yielding 140,973 labeled pairs that include contextual cues. We further validate that, by providing comprehensive semantic cues absent in traditional payload-only corpora, this context-rich dataset acts as the foundation for enhancing model performance and enables more accurate identification of both known and emerging SQLi attacks. This work emphasizes that the request-response context is not merely an auxiliary feature but a core enabler of effective SQLi detection, and that a well-constructed context dataset is critical for bridging the gap between existing defenses and the demands of real-world attack scenarios.

\section{Related work}
\label{sec:format}


\subsection{Traditional Methods}
Traditional SQL injection defenses mainly rely on input validation, parameterized queries, the principle of least privilege, and input filtering. Input validation and filtering restrict malicious characters and malformed inputs, while parameterized queries bind user inputs to prevent query manipulation \cite{alshammari2023deep, abdullayev2023sql, ibarra2021effective}. The principle of least privilege further limits potential damage by restricting database access \cite{jero2021practical}. In addition, lexical and syntactic analysis methods detect SQL injection by parsing query structures and patterns \cite{mehta2023sqliml}. Although these approaches are straightforward to deploy, they typically depend on predefined rules or signatures and require frequent updates to handle evolving attack techniques.

\subsection{Machine Learning Methods}
Recent studies have explored machine learning and deep learning techniques for SQL injection detection. Early approaches formulate SQLi detection as a classification problem based on handcrafted features extracted from query strings \cite{arasteh2024detecting}. More recently, pre-trained language models such as BERT have been adopted to capture deeper semantic representations of SQL queries, achieving improved detection performance \cite{ liu2023research}. BERT-based semantic embedding and scoring methods further improve the discrimination between benign and malicious queries \cite{sooksatra2023attribution, lu2023semantic}. Despite these advances, most existing methods focus on isolated query payloads and largely overlook the broader contextual signals available in real-world systems.

To address similar limitations in other security domains, recent work has explored multi-agent collaborative frameworks that integrate complementary contextual signals. For example, Argus demonstrates that multi-agent collaboration can effectively reduce false positives in complex security detection tasks, motivating the use of multi-agent systems to move beyond single-view analysis \cite{wang2025argus}.


\section{Dataset Building Framework}
\label{sec:Data-Collection}
\begin{figure*}
\centering
\includegraphics[width=\linewidth]{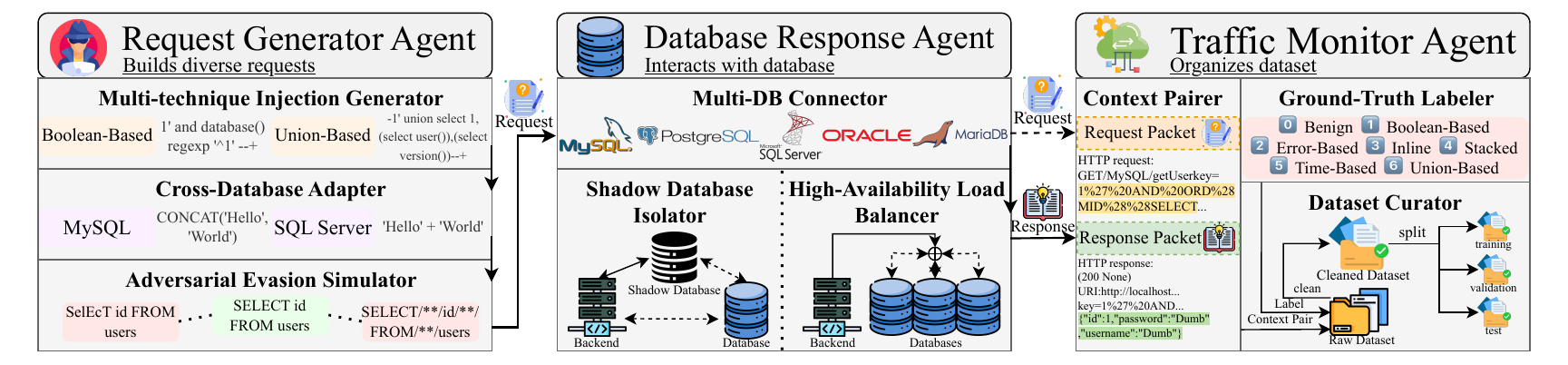}
\caption{This graph shows the framework of data collecting procedure.}
\label{fig:data_collection}
\end{figure*}

To construct a contextualized SQL-injection dataset suitable for training and evaluation, we designed a multi-agent honeypot framework, as shown in Fig.\ref{fig:data_collection}. The framework decomposes the data collection process into three goal-oriented agents that collaborate through autonomous decision-making and feedback-driven interaction, rather than static functional modules. The agents jointly form a closed-loop process in which request generation, database execution, and data quality control iteratively influence each other, enabling adaptive optimization of traffic realism and labeling consistency.

\subsection{Request Generator Agent}
The Request Generator Agent produces diverse malicious and benign HTTP requests. Rather than relying on fixed templates, it constructs attack payloads based on OWASP Top 10 SQLi categories, recent CVE vulnerabilities, and patterns observed in mainstream penetration testing tools. It adapts payload syntax via Database Response Agent feedback (Section~\ref{sec:db_agent}) and optimizes evasion strategies based on shadow database signals, replicating attackers’ trial-and-error behavior. To support this adaptive generation process, the agent is equipped with three dedicated tools:

\begin{itemize}
\item \textbf{Multi-technique Injection Generator}: Preconfigured with templates covering all common SQL injection techniques. The agent invokes this tool to generate targeted requests, ensuring coverage of diverse injection modes without over-reliance on a single type.
\item \textbf{Cross-Database Adapter}: Stores syntax rules for major relational databases including MySQL, PostgreSQL, Oracle, SQLite, SQL Server, and MariaDB. The agent uses this tool to adjust payload syntax, enabling compatibility with different database environments.
\item \textbf{Adversarial Evasion Simulator}: Integrates common obfuscation, encoding, and semantic variation techniques. The agent calls this tool to introduce modifications to base payloads, mimicking the evasion strategies employed by real attackers.
\end{itemize}

All generated requests $X_{Req}$ are passed to the Database Response Agent together with their associated injection intent for downstream execution.

\subsection{Database Response Agent}
\label{sec:db_agent}
The Database Response Agent mediates interactions between generated requests and backend databases. Under the injection intent specified upstream, this agent makes execution-level decisions through its associated tools, shaping how requests are processed and how response behaviors emerge.

\begin{itemize}
\item \textbf{Multi-DB Connector}: 
Integrates native drivers for various Relational Database Management Systems. The agent uses this tool to establish connections with target databases, ensuring that the database responses are authentic and consistent with real-system behavior.
\item \textbf{Shadow Database Isolator}: Implements routing rules that forward all requests to cloned "shadow databases" (with identical structures to production databases) instead of real business databases. This tool protects production data from destructive queries while preserving genuine database feedback.
\item \textbf{High-Availability Load Balancer}: Manages a pool of shadow database replicas. The agent invokes this tool to distribute incoming requests across replicas, preventing service disruption under high attack loads and ensuring stable traffic processing.
\end{itemize}

After processing each request, the agent provides the request $X_{Req}$ with its database response $X_{Resp}$ to the Traffic Monitor Agent for downstream interpretation.

\subsection{Traffic Monitor Agent }
The Traffic Monitor Agent transforms raw traffic into structured, ready data by interpreting observed responses imposed by upstream agents through sequentially invoking the following tools.

\begin{itemize}
\item \textbf{Context Pairer $T^{{Mon}}_1$}: Uses metadata, such as session IDs and timestamps, to pair HTTP requests from the Request Generator Agent with corresponding server responses from the Database Response Agent, eliminating mismatches between requests and responses.
Formally, the Context Pairer outputs a request-response pair as
\begin{equation}
    X_{Pair} = T^{{Mon}}_1(X_{Req}, X_{Resp})
\end{equation}

\item \textbf{Ground-Truth Labeler $T^{{Mon}}_2$}: Maps injection techniques from the tool configurations of the Request Generator Agent to integer labels. The agent uses this tool to assign accurate labels to paired records, ensuring alignment with real attack categories. Formally, the Context Pairer outputs a label as
\begin{equation}
    Y = T^{{Mon}}_2(X_{Pair}), \quad y \in \{0,1,\dots,6\}
\end{equation}

\item \textbf{Dataset Curator $T^{{Mon}}_3$}: A data processing module dedicated to refining labeled request-response pairs into a high-quality dataset. This tool performs data cleaning by removing corrupted records (e.g., incomplete responses caused by network interruptions) and duplicate entries, both of which help eliminate noise.
It also conducts stratified partitioning, splitting data into training, validation, and test subsets while preserving the distribution of attack categories to avoid sampling bias. Formally, the Dataset Curator outputs a dataset as  
\begin{equation}
    D = T^{{Mon}}_3(X_{pair}, Y)
\end{equation}
\end{itemize}

In conclusion, the Traffic Monitor Agent produces a request element:  
\begin{equation}
D = T^{Mon}_3\!\left(T^{Mon}_2\!\left(T^{Mon}_1(X_{Req}, X_{Resp})\right)\right)
\end{equation}

This process yields a semantically consistent request-response dataset.

\subsection{Request-Response Context Dataset }
After data cleaning, the framework produced 140,973 request-response pairs in total, each annotated with an integer label (0–6) denoting injection type (benign, boolean-based, error-based, inline, stacked, time-based blind, or union-based). Each record contains:
\begin{itemize}
    \item \textbf{Request-Response Pair}: Combined request and matching server responses, retaining semantic hints about attack success, since responses frequently contain indicators such as error logs, response delays, or status codes that are absent in payload-only datasets.
    \item \textbf{Label}: Based on the injected technique. 
\end{itemize}

 This dataset focuses on the importance of bidirectional context. Such contextual features significantly enhance model generalization, as validated in our subsequent experiments. 

\section{Evaluation and Analysis}



\subsection{Model Selection}
In real-world deployment scenarios, SQL injection detection systems must operate under strict latency constraints, which imposes significant performance requirements on detection models \cite{neupane2025detecting}. While large-scale neural networks deliver high detection accuracy, their substantial computational and memory overhead renders them impractical for real-time traffic inspection on commodity hardware or edge devices. Lightweight neural network architectures are therefore essential to balance throughput and responsiveness—for this reason, we first evaluate four representative lightweight models: CNN (Convolutional Neural Network), RNN (Recurrent Neural Network), LSTM (Long Short-Term Memory), and BiLSTM (Bidirectional Long Short-Term Memory). These models are selected for their widespread application in sequence data processing and their inherent efficiency, making them well-suited for resource-constrained deployment environments while still retaining basic capabilities for pattern recognition in SQLi-related data. 

\subsection{Datasets}

To validate the role of request-response context in improving SQL injection detection performance, we use two types of datasets with different color notations: the Context dataset is marked as \colorbox{mygreen}{\phantom{X}}, and the Payload dataset as \colorbox{myred}{\phantom{X}}. It is crucial to emphasize that these two datasets are homologous: both are derived from the original traffic generated by the multi-agent honeypot framework detailed in Section \ref{sec:Data-Collection}. They share an identical data distribution, covering the same categories of benign requests and SQLi attack types. The only distinction lies in their data collection scope: the Context dataset captures complete bidirectional request-response context traffic, including request payloads, server responses with and associated metadata; in contrast, the Payload dataset only extracts the request payload portion from the same original traffic, excluding all response-related context information. These two datasets are applied in two classification tasks for SQL injection detection: the 2-class task adopts a binary division logic, where label 0 (representing benign traffic) is treated as one category, and labels 1–6 (corresponding to all malicious attack types) are merged into a single category of "malicious traffic" to distinguish only between benign and malicious traffic; in contrast, the 7-class task treats each of the labels 0–6 as an independent category, further classifying malicious traffic into the six specific attack types mentioned above.

\subsection{Performance of Context Dataset on Lightweight Models}

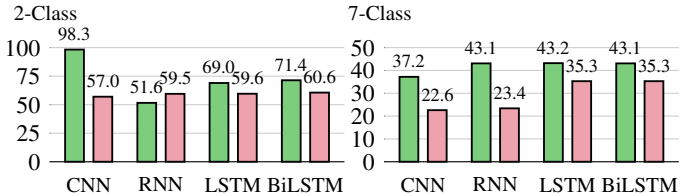
\begin{figure}[h]
    \hspace*{-0.5cm}
    \begin{tikzpicture}
          \centering
          \begin{groupplot}[
            height=3.1cm,
            /pgf/bar width=0.25cm,
            axis x line*=bottom, axis y line*=left, enlarge x limits=true,
            xticklabel style={yshift=-0.8mm, font=\small, align=center},
            ybar=3.6pt, clip=false,
            ymin=0, ymax=100, ytick={0, 25, 50, 75, 100}, yticklabels={0, 25, 50, 75, 100},
            ymajorgrids, major grid style={draw=black!20}, tick align=inside,
            yticklabel style={font=\small}, tickwidth=0pt,
            y axis line style={opacity=0},
            group style={group size=4 by 2, horizontal sep=15pt, vertical sep=33pt},
          ]
      
            \nextgroupplot[ymin=0, ymax=100, ytick={0, 25, 50, 75, 100}, yticklabels={0, 25, 50, 75, 100}, ylabel={\footnotesize 2-Class}, y label style={at={(0.3, 1.3)}, rotate=-90}, xticklabels={CNN, RNN, LSTM, BiLSTM}, xmin=-0.2, xmax=3.2, xtick={0, 1, 2, 3}, width=5.5cm, xticklabel style={font=\footnotesize}]
            
              \addplot [line width=0.7pt, fill=mygreen, error bars/.cd, y dir=both, y explicit, error bar style={draw=black}] coordinates {
                (0, 98.3)
              (1, 51.6)
              (2, 69.0)
              (3, 71.4)
              };
              \node[above] at ($(axis cs:-0.2, 98.3)$) {\scriptsize 98.3};
            \node[above] at ($(axis cs:0.8, 51.6)$) {\scriptsize 51.6};
            \node[above] at ($(axis cs:1.8, 69.0)$) {\scriptsize 69.0};
            \node[above] at ($(axis cs:2.8, 71.4)$) {\scriptsize 71.4};        
    
              \addplot [line width=0.7pt, fill=myred, error bars/.cd, y dir=both, y explicit, error bar style={draw=black}] 
              coordinates {
                (0, 57.0)
              (1, 59.5)
              (2, 59.6)
              (3, 60.6)
              };
              \node[above] at ($(axis cs:0.2, 57.0)$) {\scriptsize 57.0};
            \node[above] at ($(axis cs:1.2, 59.5)$) {\scriptsize 59.5};
            \node[above] at ($(axis cs:2.2, 59.6)$) {\scriptsize 59.6};
            \node[above] at ($(axis cs:3.2, 60.6)$) {\scriptsize 60.6};  
    
            \nextgroupplot[ymin=0, ymax=50, ytick={0, 10, 20, 30, 40, 50}, yticklabels={0, 10, 20, 30, 40, 50}, ylabel={\footnotesize 7-Class}, y label style={at={(0.3, 1.3)}, rotate=-90}, xticklabels={CNN, RNN, LSTM, BiLSTM}, xmin=-0.2, xmax=3.2, xtick={0, 1, 2, 3}, width=5.5cm, xticklabel style={font=\footnotesize}]
            
              \addplot [line width=0.7pt, fill=mygreen, error bars/.cd, y dir=both, y explicit, error bar style={draw=black}] coordinates {
                (0, 37.2)
              (1, 43.1)
              (2, 43.2)
              (3, 43.1)
              };
              \node[above] at ($(axis cs:-0.2, 37.2)$) {\scriptsize 37.2};
            \node[above] at ($(axis cs:0.8, 43.1)$) {\scriptsize 43.1};
            \node[above] at ($(axis cs:1.8, 43.2)$) {\scriptsize 43.2};
            \node[above] at ($(axis cs:2.8, 43.1)$) {\scriptsize 43.1};        
    
              \addplot [line width=0.7pt, fill=myred, error bars/.cd, y dir=both, y explicit, error bar style={draw=black}] 
              coordinates {
                (0, 22.6)
              (1, 23.4)
              (2, 35.3)
              (3, 35.3)
              };
              \node[above] at ($(axis cs:0.2, 22.6)$) {\scriptsize 22.6};
            \node[above] at ($(axis cs:1.2, 23.4)$) {\scriptsize 23.4};
            \node[above] at ($(axis cs:2.2, 35.3)$) {\scriptsize 35.3};
            \node[above] at ($(axis cs:3.2, 35.3)$) {\scriptsize 35.3};  
    
          \end{groupplot}
        \end{tikzpicture}
    \caption{Accuracy comparison of traditional models on Payload vs. Context datasets}
    \label{fig:trad_model_ctx}
\end{figure}

As shown in Fig.\ref{fig:trad_model_ctx}, models trained on the Context dataset generally tend to outperform those trained on the Payload dataset in most cases, either by large or minor margins, only the RNN has a slight deviation in the 2-class classification task. However, this situation is a rare edge case and does not change the overall trend of the Context dataset improving the model's SQL injection detection performance. Specifically, CNN
showed a 41.3\% improvement in 2-class classification task and a 14.6\% improvement in 7-class classification task, and BiLSTM improved by 10.8\% in 2-class and 7.8\% in 7-class. 

This trend holds in both 2-class and 7-class settings, demonstrating that contextual features from server responses provide crucial signals—such as error codes, response delays, or anomalous outputs—that are invisible in payload-only data.

However, the absolute accuracy of lightweight models on the Context dataset remains unsatisfactory. Even with contextual information, limited-capacity models struggle to fully capture the complexity of SQLi patterns. To further validate whether the Context dataset can continue to exert its advantages in more advanced experimental settings, we designed additional experiments, focused on verifying the effectiveness of the Context dataset on knowledge distillation.

\subsection{Performance of Context Dataset on Knowledge Distillation}
Knowledge distillation is a standard technique for improving the accuracy of small-scale models, which transfers knowledge from a high-capacity teacher model to a lightweight student model. To address the unsatisfactory absolute accuracy of lightweight models observed in previous evaluations and further verify the effectiveness of the Context dataset in advanced model optimization scenarios, we conduct evaluations on lightweight models enhanced by knowledge distillation technique. Specifically, we assess the performance of distilled lightweight student models across different architectures (CNN, RNN, LSTM, BiLSTM) under both 7-class and 2-class SQLi classification tasks.



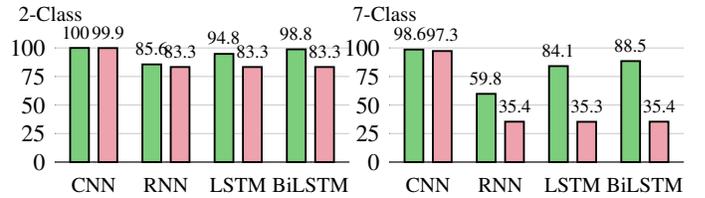
\begin{figure}[h]
    \hspace*{-0.5cm}
    \begin{tikzpicture}
          \centering
          \begin{groupplot}[
            height=3.1cm,
            /pgf/bar width=0.25cm,
            axis x line*=bottom, axis y line*=left, enlarge x limits=true,
            xticklabel style={yshift=-0.8mm, font=\small, align=center},
            ybar=3.6pt, clip=false,
            ymin=0, ymax=100, ytick={0, 25, 50, 75, 100}, yticklabels={0, 25, 50, 75, 100},
            ymajorgrids, major grid style={draw=black!20}, tick align=inside,
            yticklabel style={font=\small}, tickwidth=0pt,
            y axis line style={opacity=0},
            group style={group size=4 by 2, horizontal sep=15pt, vertical sep=33pt},
          ]
      
            \nextgroupplot[ymin=0, ymax=100, ytick={0, 25, 50, 75, 100}, yticklabels={0, 25, 50, 75, 100}, ylabel={\footnotesize 2-Class}, y label style={at={(0.3, 1.3)}, rotate=-90}, xticklabels={CNN, RNN, LSTM, BiLSTM}, xmin=-0.2, xmax=3.2, xtick={0, 1, 2, 3}, width=5.5cm, xticklabel style={font=\footnotesize}]
            
              \addplot [line width=0.7pt, fill=mygreen, error bars/.cd, y dir=both, y explicit, error bar style={draw=black}] coordinates {
                (0, 100)
              (1, 85.6)
              (2, 94.8)
              (3, 98.8)
              };
              \node[above] at ($(axis cs:-0.25, 100)$) {\scriptsize 100};
            \node[above] at ($(axis cs:0.8, 85.6)$) {\scriptsize 85.6};
            \node[above] at ($(axis cs:1.8, 94.8)$) {\scriptsize 94.8};
            \node[above] at ($(axis cs:2.8, 98.8)$) {\scriptsize 98.8};        
    
              \addplot [line width=0.7pt, fill=myred, error bars/.cd, y dir=both, y explicit, error bar style={draw=black}] 
              coordinates {
                (0, 99.9)
              (1, 83.3)
              (2, 83.3)
              (3, 83.3)
              };
              \node[above] at ($(axis cs:0.2, 99.9)$) {\scriptsize 99.9};
            \node[above] at ($(axis cs:1.2, 83.3)$) {\scriptsize 83.3};
            \node[above] at ($(axis cs:2.2, 83.3)$) {\scriptsize 83.3};
            \node[above] at ($(axis cs:3.2, 83.3)$) {\scriptsize 83.3};  
    
            \nextgroupplot[ymin=0, ymax=100, ytick={0, 25, 50, 75, 100}, yticklabels={0, 25, 50, 75, 100}, ylabel={\footnotesize 7-Class}, y label style={at={(0.3, 1.3)}, rotate=-90}, xticklabels={CNN, RNN, LSTM, BiLSTM}, xmin=-0.2, xmax=3.2, xtick={0, 1, 2, 3}, width=5.5cm, xticklabel style={font=\footnotesize}]
            
              \addplot [line width=0.7pt, fill=mygreen, error bars/.cd, y dir=both, y explicit, error bar style={draw=black}] coordinates {
                (0, 98.6)
              (1, 59.8)
              (2, 84.1)
              (3, 88.5)
              };
              \node[above] at ($(axis cs:-0.25, 98.6)$) {\scriptsize 98.6};
            \node[above] at ($(axis cs:0.8, 59.8)$) {\scriptsize 59.8};
            \node[above] at ($(axis cs:1.8, 84.1)$) {\scriptsize 84.1};
            \node[above] at ($(axis cs:2.8, 88.5)$) {\scriptsize 88.5};        
    
              \addplot [line width=0.7pt, fill=myred, error bars/.cd, y dir=both, y explicit, error bar style={draw=black}] 
              coordinates {
                (0, 97.3)
              (1, 35.4)
              (2, 35.3)
              (3, 35.4)
              };
              \node[above] at ($(axis cs:0.2, 97.3)$) {\scriptsize 97.3};
            \node[above] at ($(axis cs:1.2, 35.4)$) {\scriptsize 35.4};
            \node[above] at ($(axis cs:2.2, 35.3)$) {\scriptsize 35.3};
            \node[above] at ($(axis cs:3.2, 35.4)$) {\scriptsize 35.4};  
    
          \end{groupplot}
        \end{tikzpicture}
    \caption{Accuracy comparison of knowledge distilled models on Payload vs. Context datasets}
    \label{fig:time/acc}
\end{figure}

As shown in Fig.\ref{fig:time/acc}, on the distilled scenario, models trained on the Context dataset also significantly outperform those trained on the Payload dataset. This advantage is particularly notable in the 7-class classification task, where BiLSTM achieved a 53.1\% improvement, LSTM a 48.8\% improvement, and RNN a 24.4\% improvement.



\section{Conclusion}
This paper proposed a context-enriched SQL injection detection framework based on a multi-agent honeypot system. The Request Generator Agent, Database Response Agent, and Traffic Monitor Agent collaboratively construct a request-response context dataset, capturing semantic cues absent in payload-only corpora. Experiments across multiple models demonstrate significant accuracy gains, validating that contextual data is essential for generalizable SQLi detection.

\bibliographystyle{IEEEbib}
\bibliography{refs}

\end{document}